\documentstyle[12pt]{article}        
\begin{document}
\setlength{\textwidth}{15truecm}
\setlength{\textheight}{23cm}
\baselineskip=16pt
\bibliographystyle{unsrt}
\begin{center}
{\Large{\bf {Orientational and Translational Hopping 
in Supercooled Liquids and Glasses : Correlated Dynamics in a Free Energy 
Landscape}}} \\

\vspace{1cm}
{\large{\bf Sarika Bhattacharyya, Arnab Mukherjee and Biman Bagchi\footnote[1]{For correspondence: bbagchi@sscu.iisc.ernet.in}\\}}
\vspace{0.5cm}
 Solid State and Structural Chemistry Unit,
Indian Institute of Science,
Bangalore, India 560 012.\\
\end{center}
\vspace{-0.0cm}
\baselineskip=22pt
\begin{center}
{\large \bf Abstract}
\end{center}

  Orientational relaxation (OR) in a viscous, glassy liquid is 
investigated by carrying out extensive NPT molecular dynamics simulations
 of isolated ellipsoids in a glass forming binary mixture. 
 Near the glass transition,
 the OR occurs mainly via correlated hopping, sometimes involving  
participation of several neighboring atoms, placed in a ring like tunnel. 
In the glassy state, hopping 
is found to be accompanied by larger fluctuations in the total energy and 
the volume of the system. Both orientational and translational hopping 
are found to be {\it gated}, restricted primarily by the entropic bottlenecks,
with orientation becoming increasingly slower than translation as the pressure
is increased. OR is heterogeneous, with a wide distribution of decay times.

\baselineskip=28pt
 
PACS numbers: 61.20.Lc, 64.70.Pf

\section{Introduction}

 Even after intense studies over many decades, several aspects of relaxation 
 in supercooled liquids remain ill-understood and controversial.
 It is now recognized that anomalies in
 the relaxation in deeply supercooled liquids arise from an interplay 
 between the dynamical cooperativity between different regions in the
 liquid (an idea originally introduced by Adam and Gibbs \cite{ag}) and
 the influence of the underlying free energy surface of the liquid which
 increasingly exerts its influence as the glass transition region 
 is approached \cite{sb,bob}. The essence of the dynamical cooperativity can arguably be
 described by the mode coupling theory \cite{gotze,sbacp} which, 
 however, fails to connect to the free energy landscape. There
 is still no comprehensive treatment of 
the cross-over from the dynamically cooperative region to
  the hopping dominated region.
The cross-over scenario is intimately connected with other concepts, 
like the fragility and the super-Arrhenius behavior 
of the supercooled liquid. 

  Orientational relaxation in glassy liquid has proved to be the most 
powerful  probe of dynamics. 
The dielectric relaxation study by Goldstein and Johari first revealed the 
emergence of the $\alpha$-$\beta$ bifurcation and the disappearance of the 
$\alpha$ relaxation, as the glass transition is approached from higher 
temperature\cite{jg}. The NMR studies of Ediger and 
coworkers find an emergence of the a new time scale near glass transition 
temperature \cite{ediger}. This new time scale is larger than $\alpha$ 
relaxation time. Ediger and other workers also found evidence of heterogeneity and decoupling of 
translational and rotational motions \cite{ediger}. 

Theoretical and computer simulation studies have mostly been directed
towards understanding spherical systems, like the Lennard-Jones and the hard 
spheres. In this study, we report detailed MD (constant pressure (P), 
temperature (T), and constant total number of particles (N)) 
simulations of orientational relaxation of a few isolated
ellipsoids in glass forming binary liquids.
The questions we address here are the following :(a) How are the 
orientational and the translational motions correlated ? 
(b) What is the energy surface for 
orientational hopping? (c) How local is the hopping phenomenon ? Or in another words, how relevant is the 
free energy landscape ? (d) How sensitive is rotational diffusion to molecular shape ? Simulation studies presented here hopefully provide 
answer to some of these questions. 

 Our solvent is represented by Kob-Andersen (KA) binary mixture 
\cite{kob1} which is known 
to be a good glass former and has been extensively studied \cite{sastry},
and our  solute probes are prolate ellipsoids.  
Pressure is kept constant by Andersen's piston method \cite{andersen} 
while in the case of temperature, a damped oscillator method has been 
adopted which keeps temperature constant at 
each and every time step \cite{brown}. The piston mass involved here is 
$0.0027(m_{A}/\sigma_{A}^4)$ which is regarded as optimum \cite{brown,haile}.
The interactions between the particles are modeled by different potentials.
The interaction between the spheres are given by Lennard-Jones Potential 
(as in KA Model) and the interaction between two 
ellipsoids with arbitrary orientations is assumed to be given by the
Gay-Berne (GB) potential \cite{gb},
\begin{eqnarray}
U_{GB} = 4\epsilon(\hat r,\hat u_1,\hat u_2)\left [\left (\frac{\sigma_{0}}
{r-\sigma(\hat r,\hat u_1,\hat u_2)+ \sigma_{o}}\right )^{12}-
\left (\frac{\sigma_{0}}{r-\sigma(\hat r,\hat u_1,\hat u_2)+ \sigma_{0}}\right )^{6}\right]
\end{eqnarray}
where $\hat u_1$ $\hat u_{2}$ are the axial vectors of the ellipsoids 1 and 2.
 $\hat r$ is the 
vector along 
the intermolecular vector $r = r_2 - r_1$, where $r_1$ and $r_2$ denote the 
centers of mass of ellipsoids 1 and 2 respectively. $\sigma(\hat r,\hat u_1,
\hat u_2)$ and 
$\epsilon(\hat r,\hat u_1,\hat u_2)$ are the orientation-dependent range and 
strength parameters
respectively. $\sigma$ and $\epsilon$ depend on the aspect ratio $\kappa$.
Finally, the interaction between a sphere and an ellipsoid is accounted for by 
a modified  GB-LJ potential given below\cite{rb1}
\begin{equation}      
	U_{Ei} = 4\epsilon_{Ei}\left [ \left(\frac{\sigma(\theta)_{Ei}}{r} 
\right )^{12} - \left (\frac{\sigma(\theta)_{Ei}}{r}\right )^6\right ]
\end{equation}
\noindent where 'E' denotes the ellipsoids and 'i' can be 'A' or 'B'. 
The expression for $\sigma(\theta)_{Ei}$ is available \cite{rb1,sbor}.

The ellipsoid in binary mixture system with the above mentioned potential is a
well behaved system and it can also exhibit the experimentally observed 
anomalous viscosity dependence of the orientational correlation time 
\cite{sbor}. Four ellipsoids were placed far from each other in a binary 
mixture of 500 particles with number of 'A' particles, $N_{A}=400$ and number
of 'B' type particles $N_{B}=100$. 
The reduced temperature is expressed as,
$T^{*}$(=$k_{B} T/\epsilon_{A}$), the reduced pressure as, 
$P^{*}$(= $P\sigma^{3}_{A}/\epsilon_{AA}$).
and the reduced density as $\rho^{*}$(=$\rho \sigma_{A}^{3}$). 
The time is scaled by $\sqrt{(m_{A}\sigma_{A}^{2}/\epsilon_{AA})}$.  
The time step of the simulation is .002 $\tau$ and the system is equilibrated 
for 1.5 $\times$ 10$^{5}$ steps. 

Systematic simulations of the Kob-Andersen model (with $T^{*}=0.8$)
have been carried out by  varying pressure  over a large 
range to study the system
from its normal liquid regime to the glassy state. 
 The dynamical nature of the system at different pressures 
can be characterized from the value of the viscosity and the nature of the 
translational hopping of all the particles and orientational hopping of the 
ellipsoids. At P$^{*}$=2.0, the system is in the normal liquid regime. 
  Here the reduced density is 1.13 and the reduced viscosity is 13.446. Here
motion is continuous and orientational correlation decays like in a normal
liquid. Both the probability distributions for spatial displacement,
 $P^{i}(r)$, and for angular displacement, 
$P(\theta)$, are Gaussian in the long time. 
This behavior changes as pressure is increased. 

At P$^{*}$=5 where $\eta^{*}$=479.5, it is  a viscous liquid.
In figure $1(a)$ we plot $P(\theta)$ at pressures 5 and 6. 
At $P^{*}=5$, the rotational motion is mostly continuous and at long time 
$P(\theta)$ is Gaussian. However, at $P^{*}=6$ viscosity $\eta$ is about 
2340. At long time $P(\theta)$ has one 
peak at small angle and there is also another peak between 160$^{\circ}$-
180$^{\circ}$  The trajectory at pressure 5 shows
continuous change where as that at pressure 6 shows sharp 180$^{\circ}$ 
rotations.
 
  Figures $1(b)$ and $1(c)$ show $P(r)$ at two different pressures, each plot
showing the same  for both the larger (A) and the smaller (B) particles.
While at $P^{*}$= 5 (not shown), both $P^{A}(r)$ and $P^{B}(r)$ are Gaussian, 
at pressure 6 the $P^{B}(r)$ is still Gaussian but $P^{A}(r)$ 
shows two 
peaks. This is because at this pressure although most of the small `B'
 particles 
still have continuous motion with some amount of hopping, the `A' particles 
move primarily through hopping. At pressure 8, both $P^{A}(r)$ and $P^{B}(r)$ show multiple peaks. At this pressure the density is around 1.36 and 
hopping is the dominant mode of transport for all the particles. 
A small peak observed in $P^{B}(r)$ at higher $r$ value ($2.5 \sigma$) 
signifies correlated hopping. 
The peak at $0.2\sigma$ describes the 
vibrational motion within the cage. {\it The cross-over from continuous to 
hopping dominated regime is 
accompanied by a dramatic slow down in the 
decay of the stress auto-time correlation function (not shown here) and the 
sharp rise in the value of the viscosity.} 
From the above analyses, we conclude that the larger (A) particles
freeze near $P^{*}$=6 while the smaller (B) particles freeze near $P^{*}$=7.

At $P^{*}=10$ the system is so densely 
 packed that there hardly remains any signature of liquid like region. 
The reduced density of this system is 1.41.
In this 
 highly dense system, each ellipsoid has about 22 neighbors. 
 Here we found primarily two different 
 kinds of translational motions, one is many particle correlated 
hopping and the other 
is motion in a ring like tunnel. 
We also found correlated orientational and translation 
hopping. Figures 2 and 3 show the displacement of the 1st and 
2nd ellipsoid in reduced time and the change in their instantaneous 
orientation. 
Study of the first ellipsoid and its neighboring particle dynamics 
reveal that about 5-6 particles 
translate (greater than 0.525 $\sigma$) during (or just prior to) the time the ellipsoid hops. 
The OCF decays considerable amount {\it only} during this period. 
Analyzing the dynamics of the second ellipsoid and its 
neighboring particles and their 
neighbors, we found that 5 particles ( including the 
2nd ellipsoid) have similar trajectories (around 1650 $\tau$). 
{\it All these particles are not the neighbor of the ellipsoid, they are 
consecutive neighbors and are placed in a ring like tunnel}. 
Thus although the ellipsoid translates considerable amount in this tunnel, 
the orientational dynamics is constrained, leading to less 
decay of the OCF ( only upto 0.82).  
We also find that hopping of the particle is 
associated with larger fluctuation in volume and total 
energy in this glassy system.  

 The study of the OCF at different pressures show that at high pressures ($P^{*}\ge$6) the 
OCF decays only through hopping and there is a distribution of relaxation 
times. While the average orientational correlation function is non-exponential,
decay of each ellipsoid is found to be only weakly non-exponential, with widely different relaxation
times, often differing by more than one order of magnitude at pressure $P^{*}$=6. 
In two different runs the average OCF is also found to be drastically 
different, although the average energy and volume of the different 
runs remain the same. Thus, the local minimum explored by each 
probe ellipsoid 
are different.

In an ordinary liquid rotational relaxation time, $\tau_{R}$ is of the 
order of $10^{-11} sec$ whereas the time taken to translate 1 molecular 
diameter, $\tau_{T}$ is 
about $10^{-10} sec$. In a supercooled liquid both translation and rotation 
slows down. From the mean square displacement we found that $\tau_{T}$ of the 
ellipsoid is of the order of $10^{-5}sec$. Since the orientational correlation function does not decay much it is difficult to make an estimate of 
$\tau_{R}$. We found that even at 1200 $\tau$ the OCF does not decay 
below 0.55. Thus it can be safely concluded that in our system 
$\tau_{2R} >> \tau_{T}$. This observation is similar to that reported by 
Ediger and co-workers where they found that the molecule translates hundreds 
of molecular diameter before it rotates once \cite{ediger}.

  Figure 4 shows the variation of the single particle potential energy 
{\it during hopping} as a function of 
distance traversed, both for the ellipsoid and two of its neighbors. 
The study of the energetics (during orientational and translation 
hopping) reveals that both kinds of hopping are {\it gated process}, 
where the 
free energy barrier (or the saddle) is {\it entropic} in nature. 
The ellipsoid hopping is 
associated with instantaneous rearrangement (significant displacement) 
of surrounding molecules. At high pressure the particles are 
densely packed and trapped in regions separated by small windows 
(entropic bottlenecks) \cite{zwanzig}. If the motion of the ellipsoid is 
in the x-direction then the surrounding particles are found to move in the 
perpendicular plane. Thus only rearrangements or displacement of surrounding molecules open up the gate and allow the ellipsoid to move as envisaged in the Zwanzig model \cite{zwanzig}. In the process of creation of the gate some of the 
particles register a small rise in energy (shown in figure 4(b) and 4(c)). 
However the sum of energies of the ellipsoid and its neighbors remains 
the same. 

The present study shows that orientational 
relaxation in a viscous liquid near glass 
transition, involves collective motion which, although complex, has certain 
regular and understandable features. For example, hopping of a particle in 
question is gated. There is clearly a waiting time distribution 
of this collective event to happen, although even very close to glass 
transition 
these events are certainly not rare. Therefore, these are expected to play 
important role in the relaxation of glassy liquids. A complete theory of these
highly collective processes would be  non-trivial, 
and would perhaps first require a
``coarse-grained'' theory of the type discussed by Xia and Wolynes\cite{xpgw}.

 We find that orientational relaxation
occurs increasingly on slower time scale, compared to translation, as the
glass transition is approached. This is because molecules can hop preserving 
its orientation and also 180$^\circ$ orientational hopping does not 
change 2nd rank
harmonics, often probed in experiments. In the glassy state the
hopping of ellipsoid is non-local. Finally, we note that the present
simulations can reproduce several behavior observed in recent experiments
on orientational relaxation in deeply supercooled liquids.

It is a pleasure to thank Prof. Peter Wolynes and Prof. Srikanth Sastry
for discussions and correspondence.
The work has been supported in parts by grants from the Council of Scientific
and Industrial Research and the Department of Science and Technology, India.

\newpage

{\large \bf Figure Captions}
   
{\bf Figure 1} (a) Dashed and Solid lines show variation of $P(\theta)$ 
against angular displacement $\theta$ for pressure $P^{*}=5.0$ and $P^{*}=6$, 
respectively. At $P^{*}=6$ the distribution is bimodal with a second peak 
appearing around $160^\circ$. (b)Distribution of $P(r)$ against 
displacement ($r^{*}$) for two different types of 
particles ( 'A' and 'B') at pressure $P^{*}=6.0$. 
Solid line and dashed line denote bigger and smaller particles, respectively.
(c) Same plot as 1(b) but at pressure $P^{*}=8.0$.

{\bf Figure 2} (a) Displacement of an ellipsoid (tag 1) occurs only by 
hopping at around $300\tau$. (b) Dot product of instantaneous orientation and 
the orientation at the time t=0 shows that orientational and translational 
hopping happen at the same time. 
(c) Decay of $C_{2R}(t)$ in different time intervals are plotted. 
The arrow points to the OCF which shows prominent decay during 
the time interval of the hopping.

{\bf Figure 3} (a) Displacement of another ellipsoid ( tag 2) where it 
undergoes motion in a ring like tunnel. (b) Dot product of the instantaneous orientation and the orientation at time t=0. Both the orientational hopping and translational motion are correlated. (c)Decay of $C_{2R}(t)$ in different time 
intervals are plotted. Although, the ellipsoid translates about 
1$\sigma$ distance, due to constrained orientational dynamics in the ring 
like tunnel the OCF decays only upto 0.82. The arrow points to the OCF, 
obtained during the period of hopping.

{\bf Figure 4} (a) Variation of the single particle potential energy (SPPE) 
of the ellipsoid with time around hopping signifies gated diffusion. (b) and 
(c) shows variation of SPPE of two  neighboring particles 
of the above ellipsoid. 

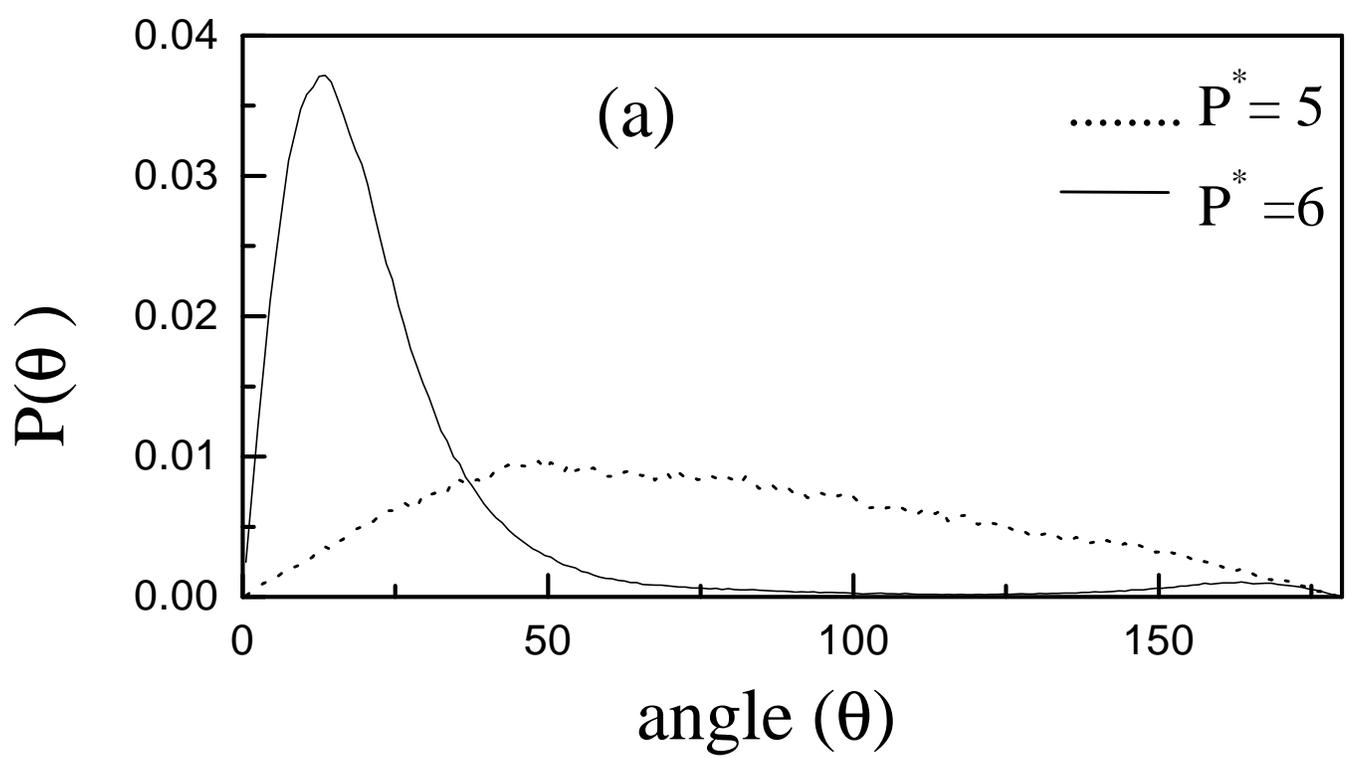
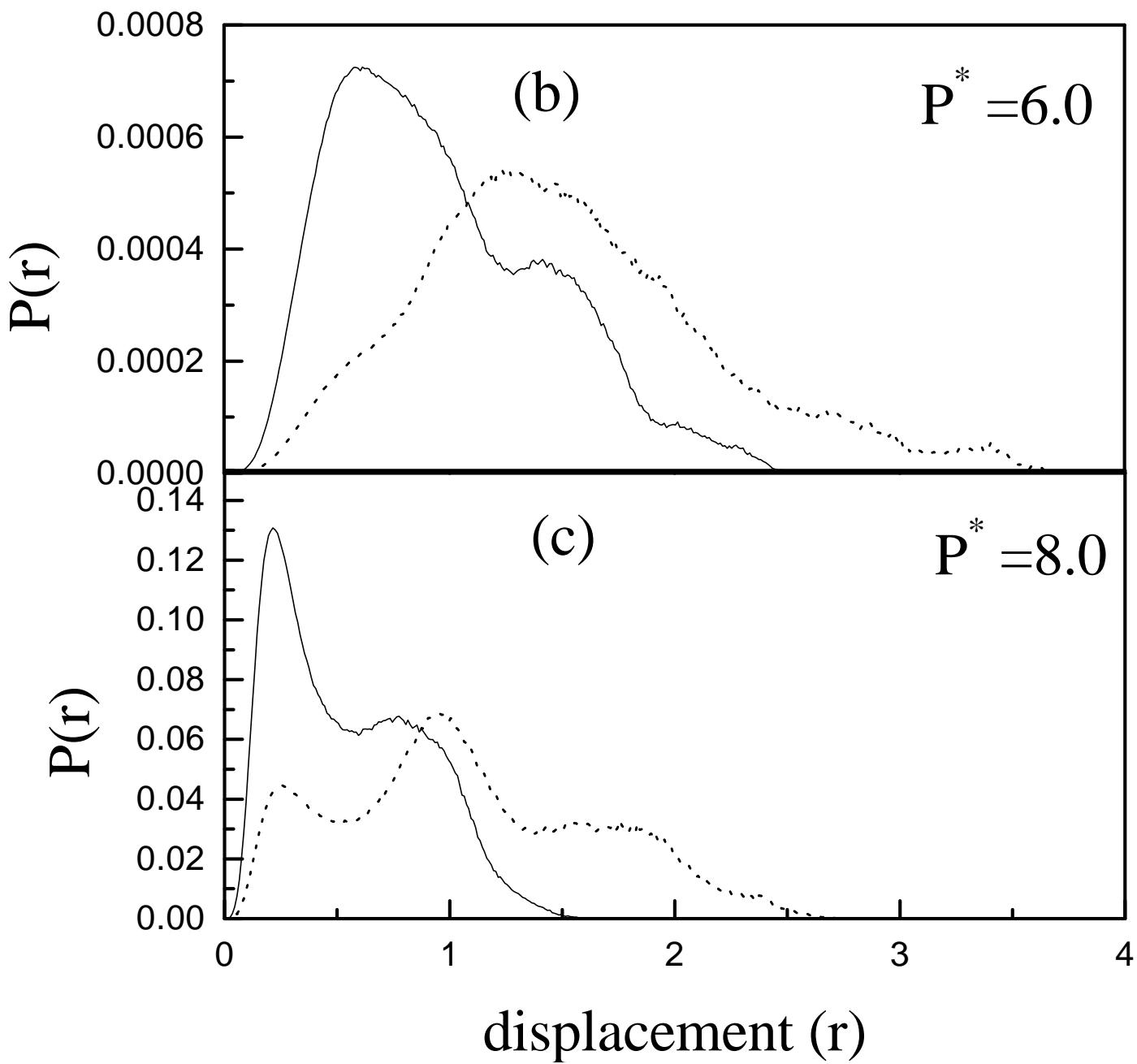
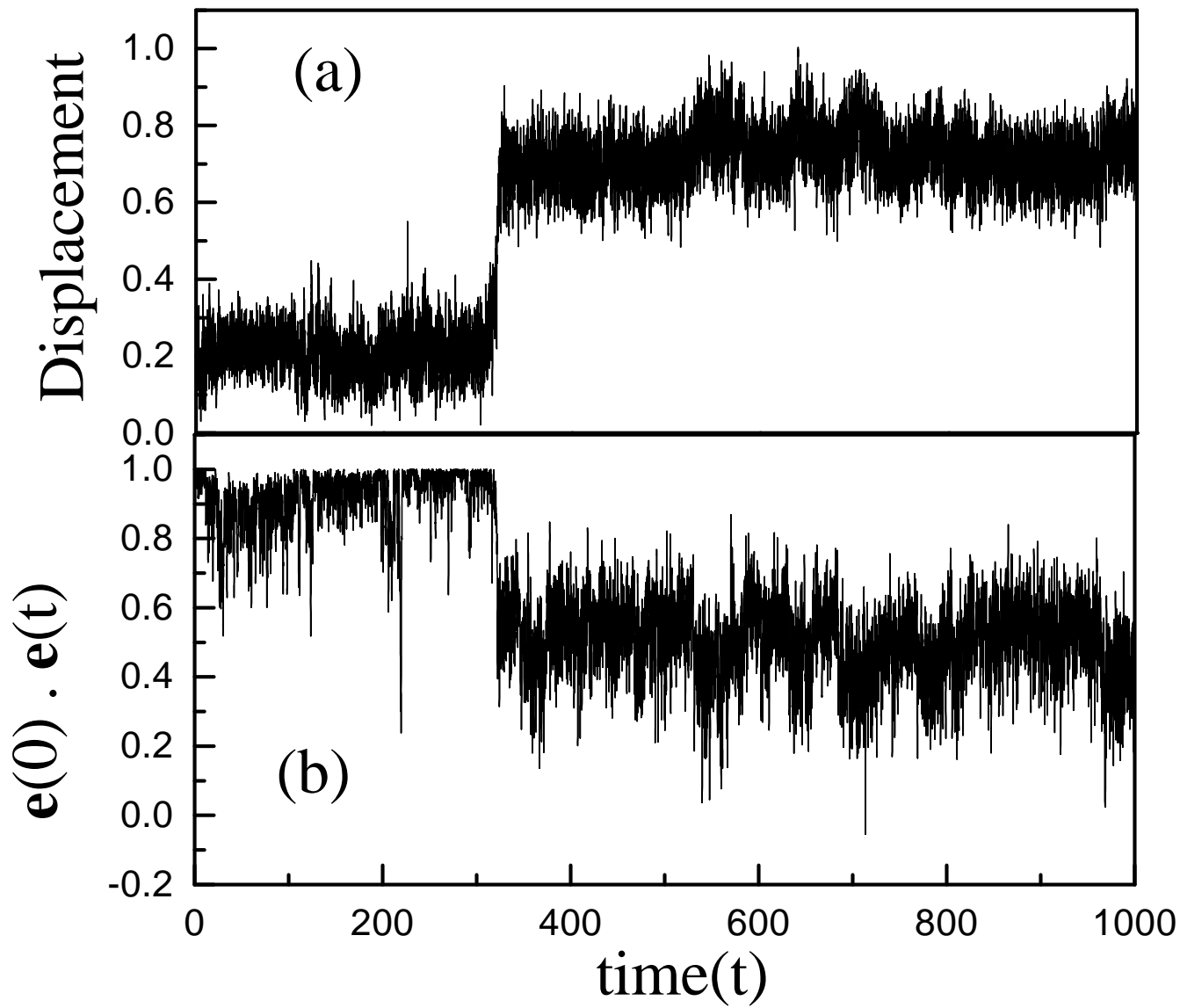
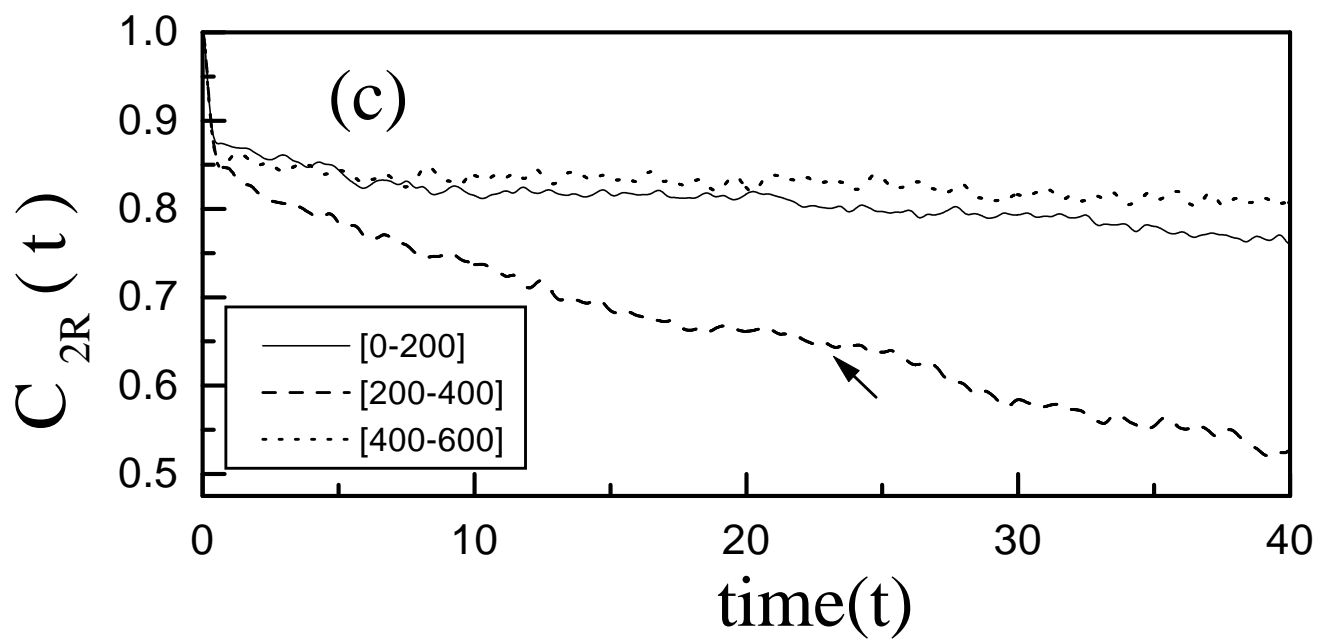
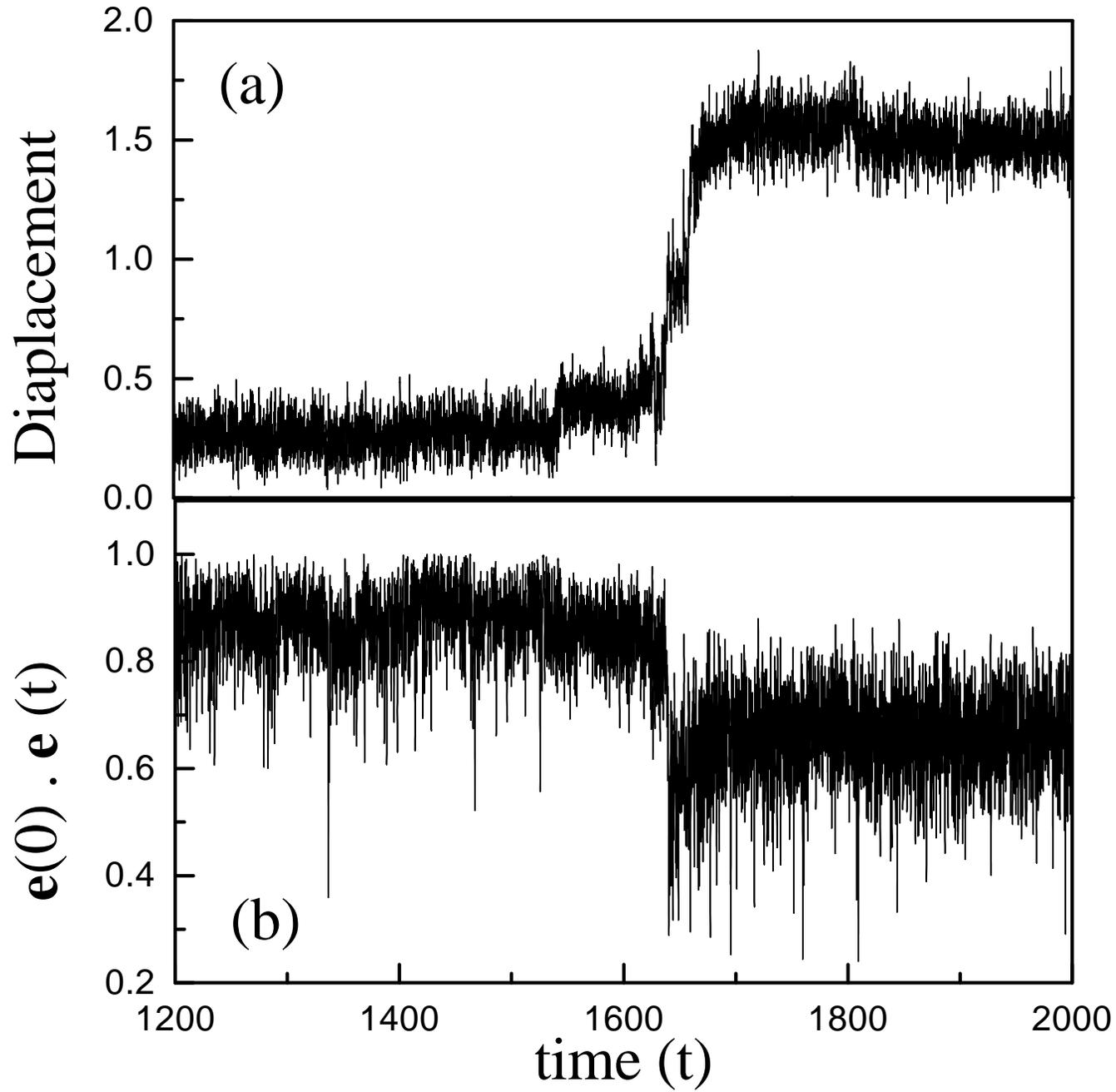
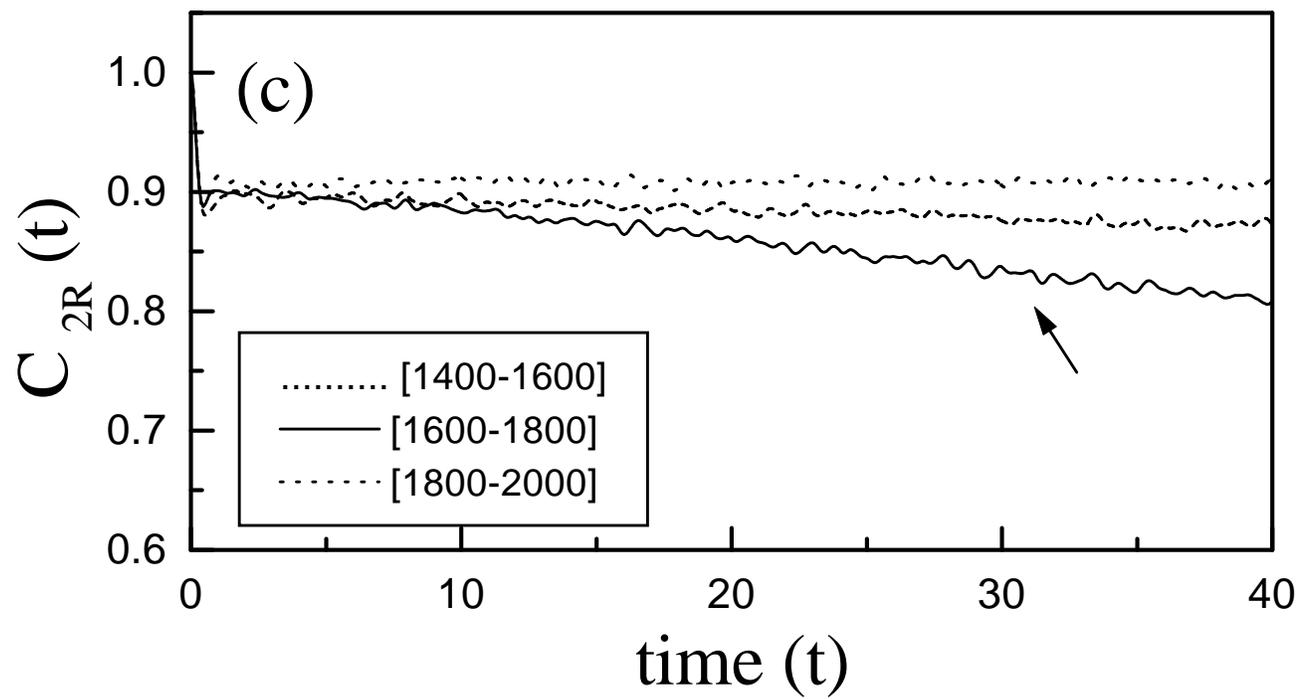
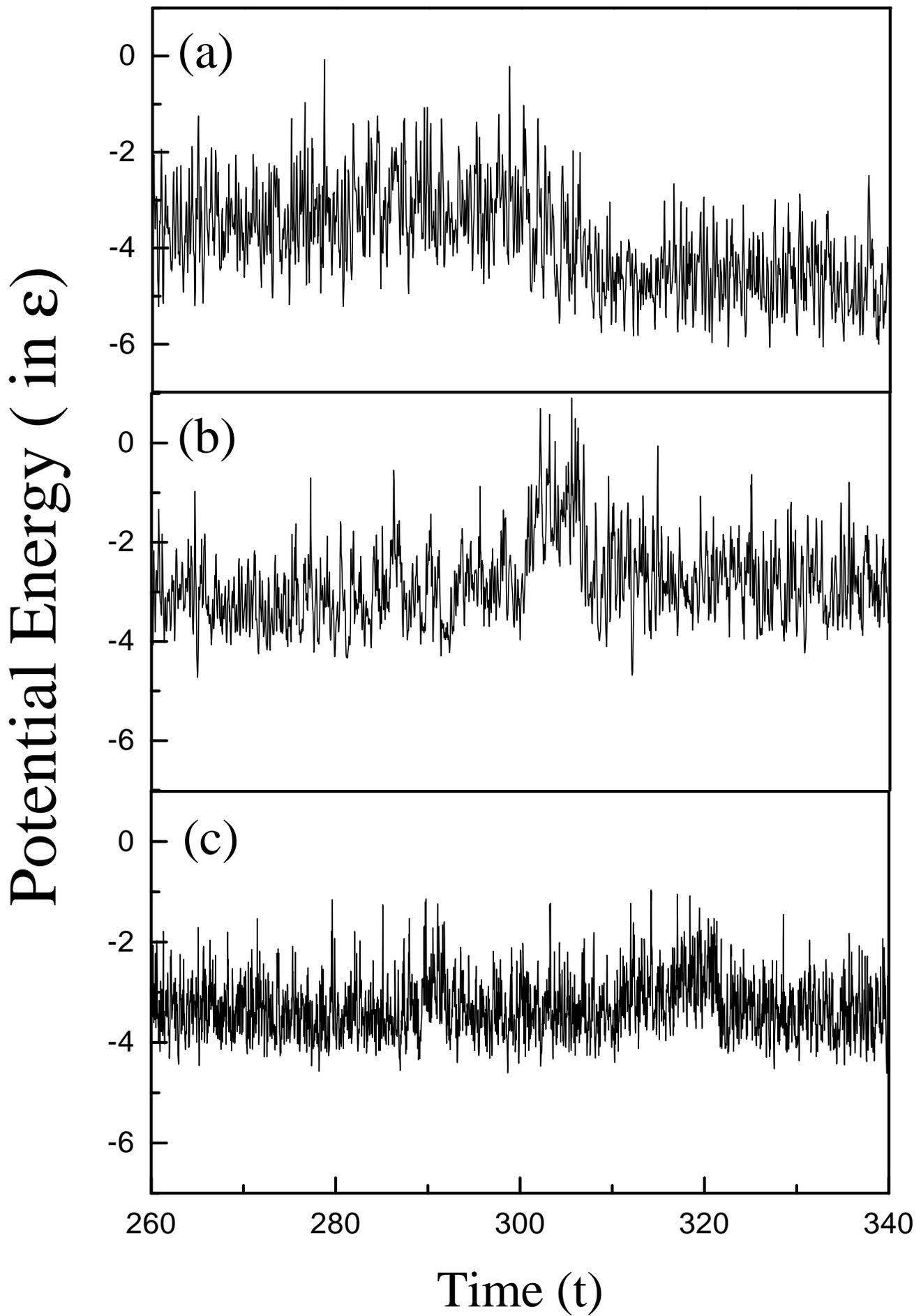
\end{document}